\documentclass[entropy,article,accept,moreauthors,pdftex,12pt,a4paper]{mdpi} 
\pdfoutput=1
\lastpage{4663}
\doinum{10.3390/e17074654}
\pubvolume{17}
\pubyear{2015}
\history{Received: 12 May 2015 / Accepted: 1 July 2015 / Published: 3 July 2015}

\usepackage{graphicx}
\usepackage{subfig}

\Title{Continuous Variable Quantum Key Distribution\\ with a Noisy Laser}

\Author{Christian S. Jacobsen, Tobias Gehring and Ulrik L. Andersen *}

\address[1]{Department of Physics, Technical University of Denmark, Fysikvej, 2800 Kongens Lyngby, Denmark; E-Mails:  chrsch@fysik.dtu.dk (C.S.J.); tobias.gehring@fysik.dtu.dk (T.G.)\vspace{-12pt}}

\corres{E-Mail: ulrik.andersen@fysik.dtu.dk.\vspace{-12pt}}

\abstract{Existing experimental implementations of continuous-variable quantum key distribution require shot-noise limited operation, achieved with shot-noise limited lasers.
However, loosening this requirement on the laser source would allow for cheaper, potentially integrated systems.
Here, we implement a theoretically proposed prepare-and-measure continuous-variable protocol and experimentally demonstrate the robustness of it against preparation noise stemming for instance from technical laser noise.
Provided that direct reconciliation techniques are used in the post-processing we show that for small distances large amounts of preparation noise can be tolerated in contrast to reverse reconciliation where the key rate quickly drops to zero.
Our experiment thereby demonstrates that quantum key distribution with non-shot-noise limited laser diodes might be feasible.}

\keyword{quantum key distribution; continuous variables; quantum information}

\PACS{03.67.Dd; 03.67.Hk; 03.67.-a}

\begin{document}


\section{Introduction}

Secure communication in the post-quantum computer era will be possible using quantum key distribution (QKD) whose security principles are based on the laws of quantum mechanics~\cite{Nielsen2000, Braunstein2005,Scarani2009,Gisin2002,Weedbrook2012a, Renner2005}.
The QKD systems of today, both proof-of-principle laboratory implementations and commercial systems, require expensive low-noise lasers and detectors. Especially QKD systems using continuous variables (CV) suffer from excess noise which rapidly decreases the secure key rate~\cite{Cerf2001, Grosshans2002,Grosshans2003, Garcia-Patron2009,Lance2005,Madsen2012}.

Prepare-and-measure CV protocols typically employ pure vacuum states randomly displaced in the amplitude and phase quadrature amplitudes of the electro-magnetic field \cite{Weedbrook2012a}.
The receiver either performs homodyne or heterodyne detection \cite{Grosshans2002,Lance2005}.
Homodyne detection measures a certain quadrature angle depending on the phase of the strong reference beam called the ``local oscillator'', which has to be switched between two orthogonal settings.
In contrast, heterodyne detection measures both orthogonal quadratures simultaneously, thereby eliminating the need of switching at the expense of one extra unit of shot-noise entering the system.
The classical post-processing is either based on direct or reverse reconciliation. For direct reconciliation Bob corrects his measurement outcomes to match Alice's preparation, while, as the name already implies, the reverse is true for reverse reconciliation.
In the shot-noise limited regime reverse reconciliation allows for larger distances between the two parties since it overcomes the $3$\,dB loss limit of direct reconciliation~\cite{Grosshans2002, Grosshans2003, Silberhorn2002}. Reverse reconciliation was introduced as an alternative to direct reconciliation and as a way to beat the intrinsic loss limit. It was also shown to be more efficient than direct reconciliation even for transmissions above $50 \ \%$ in the limit of zero excess noise~\cite{Grosshans2003a}.

Using a shot-noise limited laser, homodyne detection and reverse reconciliation a distance between the two communicating parties of $80$\,km has already been achieved~\cite{Jouguet2013}. While reaching long distances is of great importance for linking neighbouring cities, bridging short distances potentially allows for secure communication of hand-held devices with terminals. However, for hand-held devices the optical components of a QKD system have to be integrated into a chip whose manufacturing costs have to be low. To enable this, relaxing the stringent requirements on the noise of the laser becomes crucial. 

CV QKD using a noisy laser was first suggested in \cite{Filip2008, Usenko2010} using reverse reconciliation, however, it was realized that the preparation noise dramatically reduced the secure key rate, eventually making it impossible to obtain a secure key.
To regain security purification of the state was proposed by attenuating the modulated noisy state at Alice's private trusted station.
In \cite{Weedbrook2010} and \cite{Weedbrook2012} it was then suggested that short distance protocols could in fact benefit from using direct reconciliation. Surprisingly it was shown theoretically that the secure key rate for an error reconciliation efficiency of $100\%$ does not decline to zero even for arbitrary amounts of laser noise as long as the excess noise of the channel is lower than the noise of the laser.

Here, we verify this theoretical prediction by using a noisy laser beam which is used for a prepare-and-measure Gaussian modulation protocol with heterodyne detection. For collective attacks in the asymptotic limit we show that preparation noise which in the case of reverse reconciliation yields a vanishing secure key rate, can in fact be tolerated with direct reconciliation. We furthermore experimentally show that for direct reconciliation an optimal amount of preparation noise exists leading to an increased tolerable optical loss in comparison to the case of no preparation noise. 

This effect of using preparation noise to obtain greater regions of security in the parameter space can be seen as the direct reconciliation equivalent of the effect reported in \cite{Garcia-Patron2009, Pirandola2009a}. As a result of using direct reconciliation, correspondingly shot-noise limited detection is required at the receiver. More generally, the security of QKD in the presence of trusted noise (either at the preparation or detection stage) can be related to the role of quantum discord in quantum cryptography~\cite{Pirandola2014a}.

Finally, we note that Ref. \cite{Weedbrook2014} proposes a protocol for two-way continuous-variable quantum key distribution also using displaced thermal states.


\section{Theory}

Though the performed experiment is an example of a prepare-and-measure protocol, where a modulated quantum state is prepared by Alice and sent to Bob, there is a theoretical equivalence between this protocol and an entanglement based protocol~\cite{Ralph2000,Reid2000,Hillery2000}, where Alice prepares an Einstein-Podolsky-Rosen (EPR) state.
In the equivalent scheme she keeps one mode of the EPR state to herself and sends the other mode to Bob.
If Alice performs a conditioning measurement using a heterodyne detector, Bob's half of the EPR state will collapse to a coherent state.
In that way Alice prepares a coherent state with a displacement depending on her measurement outcome.

If one disregards the problems that follow from finite key sizes~\cite{Renner2005, Konig2007,Leverrier2010}, one arrives at a relatively simple bound on the secure key rate
\begin{equation} \label{eq:rate}
R = \beta I(A:B) - \chi(E:X) \ ,
\end{equation}
where $\beta$ is the reconciliation efficiency \cite{Scarani2009} and $I(A:B)$ is the classical mutual information between Alice and Bob expressed through the Shannon entropy of the corresponding classical stochastic variables of the measurements~\cite{Cover2006}.
$\chi(E:X)$ is the Holevo quantity \cite{Nielsen2000}, expressed through the von Neumann entropy of the quantum state the eavesdropper shares with the honest parties.
For reverse reconciliation $\chi(E:B) = S(\hat{E}) - S(\hat{E}|B)$, and for direct reconciliation $\chi(E:A) = S(\hat{E}) - S(\hat{E}|A)$ \cite{Weedbrook2012a}.
$S(\hat{X})$ is the von Neumann entropy of the quantum state $\rho_X$.
For Gaussian states any $\rho_X$ has a corresponding covariance matrix $\Gamma_X$, and a mean value $\langle \boldsymbol{X} \rangle$, though mean values can always be displaced out without loss of generality, as they do not contribute to the information content of the state in question.
The von Neumann entropy for a Gaussian state $\rho_X$ can be shown to be given by~\cite{Weedbrook2012a}
\begin{equation}
S(\hat{X}) = \sum_i g(\nu_i),
\end{equation}
where
\begin{equation}
g(x) = \dfrac{x+1}{2} \log_2 \left(\dfrac{x+1}{2} \right) -  \dfrac{x-1}{2} \log_2 \left(\dfrac{x-1}{2} \right),
\end{equation}
and $\nu_i$ is the $i$'th value in the symplectic spectrum of $\Gamma_X$.
The symplectic spectrum is calculated by finding the absolute eigenvalues of the matrix $i \Omega \Gamma_X$, where
\begin{equation}
\boldsymbol{\Omega} = \bigoplus_{k=1}^N \boldsymbol{\omega}
\qquad , \qquad
\boldsymbol{\omega} = 
\begin{bmatrix}
0 & 1 \\
-1 & 0
\end{bmatrix},
\end{equation}
with $N$ being the number of modes described by the state $\rho_X$. The dimension of $\Gamma_X$ is $2N \times 2N$.

An attack on a continuous-variable quantum channel between Alice and Bob is usually represented by a beam splitter, controlled by the eavesdropper Eve, who is injecting one part of an EPR state into the vacant port. Eve keeps the other mode of the EPR state in a quantum memory and interferes the first mode with all the coherent states sent by Alice. Then she does a collective measurement on the memory after all quantum states have been exchanged between Alice and Bob. This collective attack is called an ``entangling-cloner attack''~\cite{Grosshans2003,Garcia-Patron2006,Navascues2006,Weedbrook2010,Weedbrook2012a} and represents the most important and realistic type of collective Gaussian attack, whose general form is discussed in  \cite{Pirandola2008}.

\begin{figure}[H]
    \centering
    \includegraphics[width=10cm]{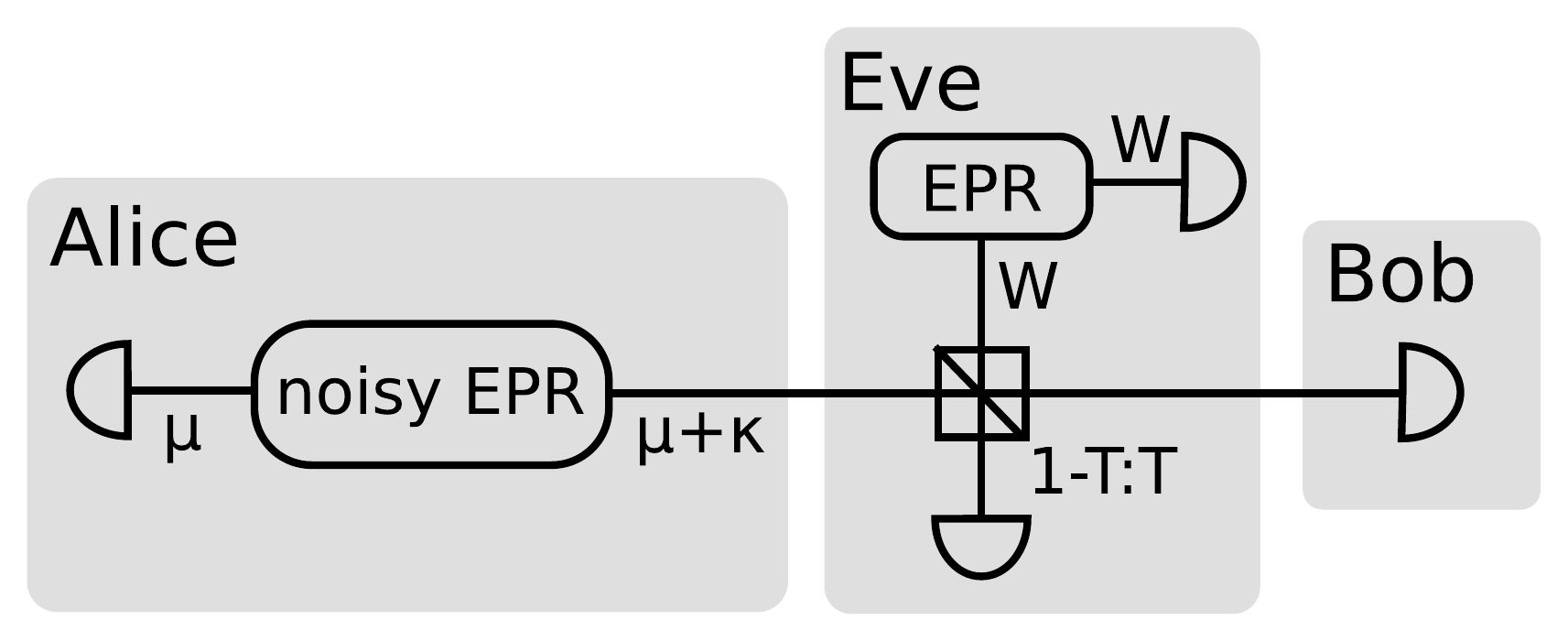}
    \caption{Equivalent entanglement based model used in the security proof. Alice produces a noisy  Einstein--Podolsky--Rosen (EPR) state which she sends to Bob. The quantum channel with transmission $T$ is controlled by the eavesdropper who injects an EPR state with variance~$W$.}
    \label{fig:entanglementbasedmodel}
\end{figure}

A schematic of the entanglement based equivalent model is shown in Figure~\ref{fig:entanglementbasedmodel}.
Alice produces a noisy EPR state
\begin{equation}
\boldsymbol{\Gamma_\text{in}} =
\begin{bmatrix}
\mu \boldsymbol{I} & \sqrt{\mu^2-1} \boldsymbol{Z} \\
\sqrt{\mu^2-1} \boldsymbol{Z} & (\mu + \kappa) \boldsymbol{I}
\end{bmatrix},
\label{eq:gamma_in}
\end{equation}
with variance $\mu$ in the mode she keeps (upper left block) and $\mu + \kappa$ in the mode she sends to Bob (lower right block). $\kappa$ represents the extra variance that comes from the preparation noise and 
\begin{equation}
\boldsymbol{I} = 
\begin{bmatrix}
1 & 0 \\
0 & 1
\end{bmatrix}
\qquad , \qquad
\boldsymbol{Z} = 
\begin{bmatrix}
0 & 1 \\
-1 & 0
\end{bmatrix}\ .
\end{equation}
Thereafter the mode sent to Bob is propagated through a beam splitter with transmittance $T$ where Eve interferes it with one mode of her EPR state. The transmitted mode is measured by Bob, while the reflected mode is measured by Eve. 
We follow the work of \cite{Weedbrook2012} where it was shown that for heterodyne detection the mutual information is given by the expression
\begin{equation} \label{eq:MutInfo}
I(A:B) = \log_2 \left(\dfrac{(1-T)W + T V_S + T V_0 + 1}{(1-T)W + TV_0 + 1} \right),
\end{equation}
where $V_S$ is the variance of the Gaussian distribution used for the signal modulation, $V_0$ is the noise carried by the light and $W$ is the variance of Eve's EPR mode.
The entanglement based protocol is related to the prepare-and-measure scheme such that $\mu = V_S + 1$.
This noise can be separated into technical and intrinsic (quantum) noise, such that $V_0 = 1 + \kappa$, in units of the shot noise.
For a protocol assuming shot-noise limited state generation $\kappa = 0$.

The symplectic spectrum of Eve's EPR state after it has been interfered with the signal mode is $\nu_{E\pm} = \dfrac{1}{2} (\sqrt{(e_V+W)^2-4T(W^2-1)} \pm (e_V-W))$, where $e_V = (1-T) V + T W$.
Similar expressions are derived in \cite{Weedbrook2012} for the symplectic spectra of the conditional states.
From these expressions of the eigenvalues one can find the Holevo information for direct and reverse reconciliation
\begin{equation} \label{eq:Holevo}
\chi(E:X) = g(\nu_{E+}) + g(\nu_{E-}) - g(\nu_{E|X+}) - g(\nu_{E|X-}),
\end{equation}
where $X$ corresponds to either conditioning on Alice or Bob depending on whether direct or reverse reconciliation is used.
The secret key rate is then easily found by combining \mbox{Equations~\eqref{eq:rate}, \eqref{eq:MutInfo} and \eqref{eq:Holevo}}.

\begin{figure}[H]
\centering
    \subfloat{\includegraphics[width=0.5\textwidth,trim=1cm 0 1cm 0,clip=true]{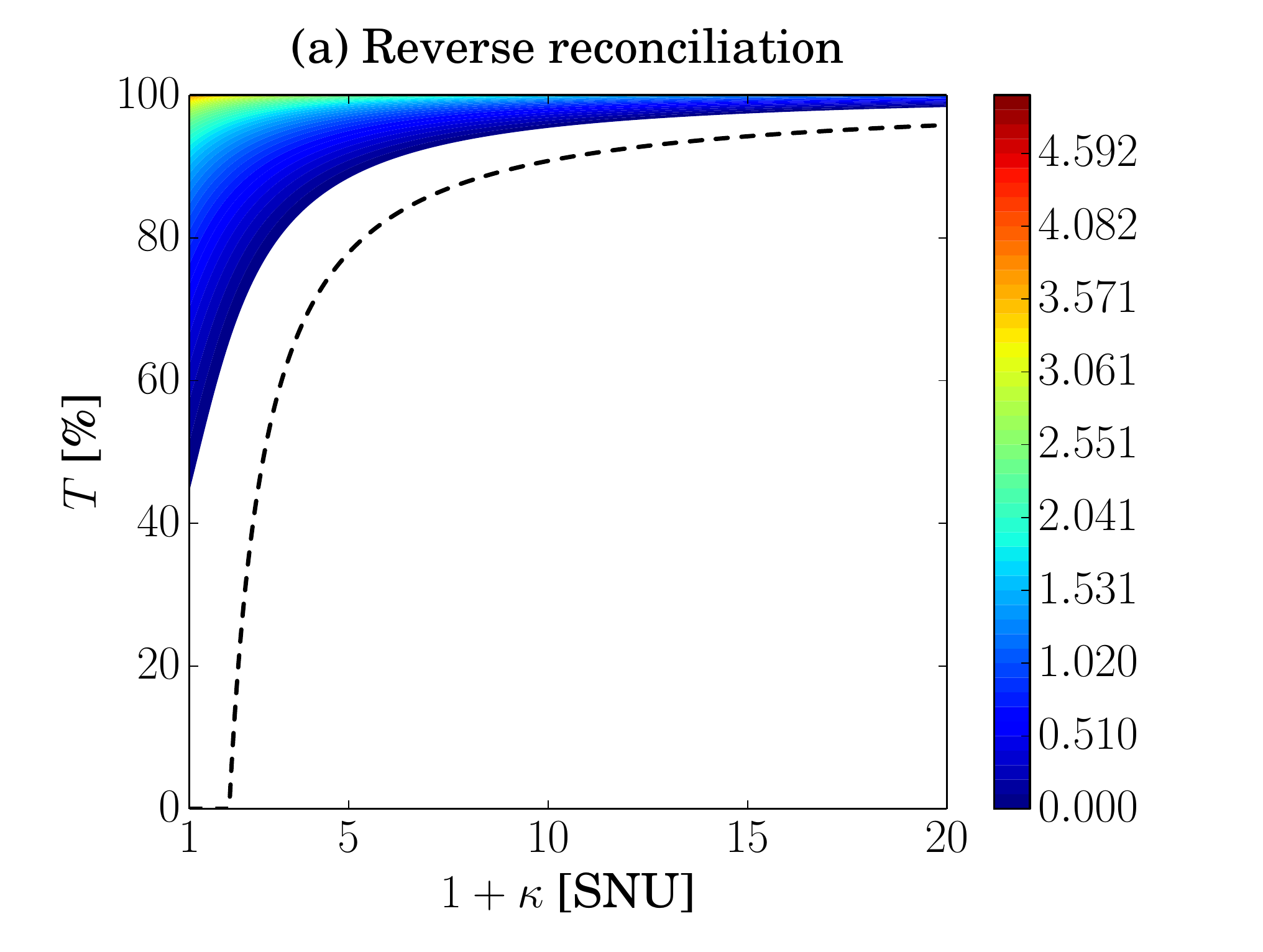}\label{fig:TheoryRR}} 
    \subfloat{\includegraphics[width=0.5\textwidth,trim=1cm 0 1cm 0,clip=true]{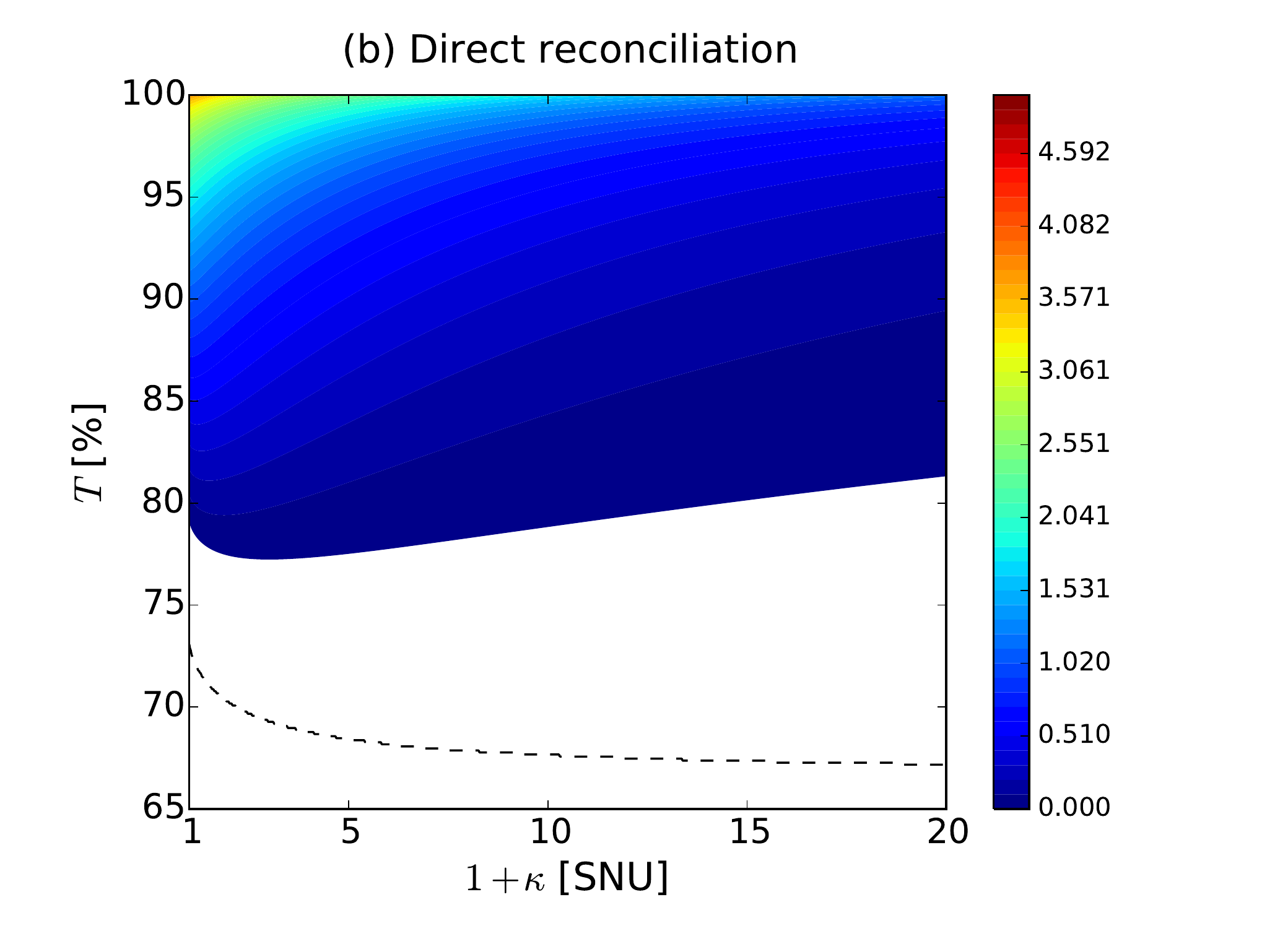}\label{fig:TheoryDR}}
    \caption{Contour plots of the secure key generation rate for varying preparation noise in shot-noise units (SNUs) and transmission $T$ for ({\bf a}) reverse reconciliation and ({\bf b}) direct reconciliation. The error reconciliation efficiency was set to $\beta = 95\%$, the modulation variance was $32$ SNUs and the channel excess noise $0.11$. The dashed lines indicate the minimal possible transmission of a channel where a positive secret key rate can still be obtained, in the ideal case for $\beta = 1$, no channel excess noise and in the limit of high modulation variance. (a) For no preparation noise ($\kappa = 0$), the rate decreases asymptotically to zero as the transmission approaches zero. When the preparation noise increases the security of reverse reconciliation is quickly compromised, to the point where almost unity transmission is required to achieve security. (b) For heterodyne detection and no preparation noise the rate goes to zero at about $79\%$ transmission, due to the extra unit of vacuum introduced by heterodyne detection. The plot shows the robustness of direct reconciliation to preparation noise.}
    \label{fig:Theory}
\end{figure}

The regions of positive key rate in terms of transmission and preparation noise are shown in Figure~\ref{fig:Theory} for both reverse and direct reconciliation.
Here one clearly sees that large preparation noise is highly detrimental to reverse reconciliation, but not to direct reconciliation which is much more robust.
In fact for direct reconciliation a preparation noise of $1+\kappa \approx 3$ is beneficial since it lowers the possible channel transmission value to about $77\,\%$ in comparison to $79\,\%$ for no preparation noise.
In the ideal case of $\beta = 1$ the secure region even keeps increasing with increasing preparation noise.
This behaviour is displayed in the figure by the dashed line calculated in the limit of high modulation variance with ideal reconciliation and no channel excess noise. 
This case was thoroughly investigated in \cite{Weedbrook2012}.
\vspace{-24pt}

\section{Experiment}

\begin{figure}[H]
\centering
\includegraphics[width=0.5\textwidth]{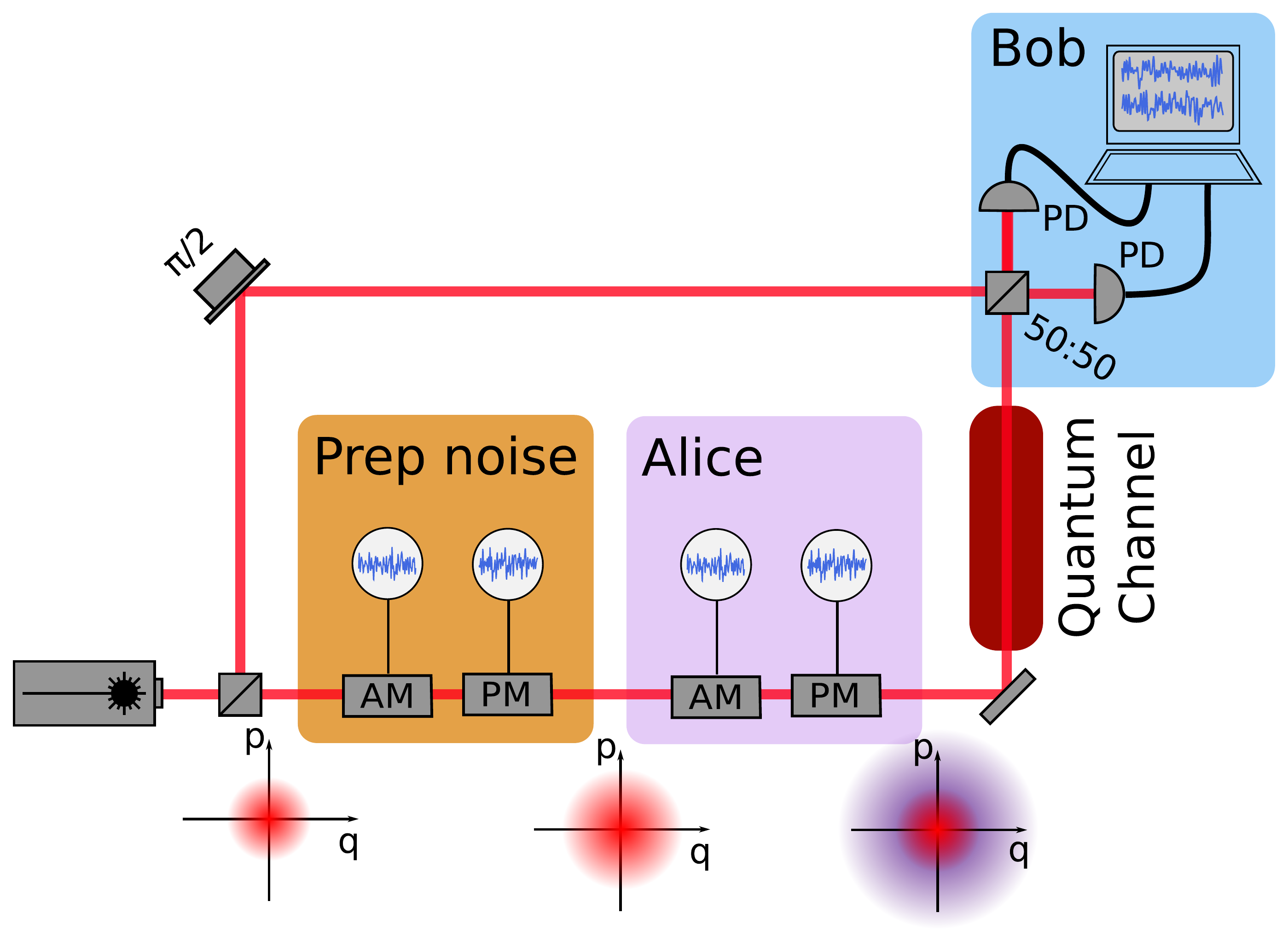}
\caption{Schematic representation of the experiment. A shot-noise limited laser is amplitude- and phase-modulated with two independent white-noise sources to simulate a noisy laser. Subsequently, Alice modulates the noisy laser beam in amplitude and phase using a known modulation and sends it to Bob through the quantum channel who performs heterodyne detection. The quantum channel's transmission was simulated by an (for coherent states) equivalent reduction of the modulation variances. AM: Amplitude Modulation. PM: Phase Modulation. PD: Photo Detector.}
\label{fig:SetupSketch}
\end{figure}

The experimental setup is shown in Figure~\ref{fig:SetupSketch}. A shot-noise limited continuous-wave laser at $1064$\,nm was used in conjunction with two electro-optical modulators and two white noise generators to simulate various degrees of preparation noise.
Since the laser was shot-noise limited a simplified heterodyne detection was performed where the signal beam is interfered with a local oscillator and locked such that the subtraction of the two photo detector signals represents the phase quadrature.
At the same time, the photo detector signals were added to measure the amplitude quadrature.
Note that typically such a detection approach would not eliminate any technical noise present in the noisy laser beam.
In a setting where the local oscillator for heterodyne detection is also derived from the noisy laser, two separate homodyne detectors measuring orthogonal quadratures have to be employed.
That way the common mode rejection ratio of the homodyne detector would provide a shot-noise limited measurement. The quantum efficiency of the photo diodes was about $90\,\%$ and the visibility at the balanced beam splitter about $93\,\%$. The optical power in the two beams was $2.8$\,mW each.

The signal modulations were delivered by two white noise generators, whose outputs were sent both to the modulators and to the data acquisition.
The acquired data from Alice's modulations and from Bob's amplitude and phase quadrature measurements represent the raw key.
To determine the secure key rate using Equation~\eqref{eq:rate} the mutual information between Alice's and Bob's data were calculated and the transmission as well as the excess noise of the channel was estimated from the data.
Transmission was estimated from a calibration measurement where the states experienced only the intrinsic loss of the setup. 
This defines unity transmission, which is possible because of the scaling of coherent states under losses.
Transmission values were calculated by relating the measured variances to this calibration measurement.
Excess noise was determined by a suitable minimization of the subtraction of the performed modulation and the measurement outcomes.
The remaining variance above shot-noise was ascribed to excess noise.

\begin{figure}[H]
\centering
    \subfloat{\includegraphics[width=0.5\textwidth]{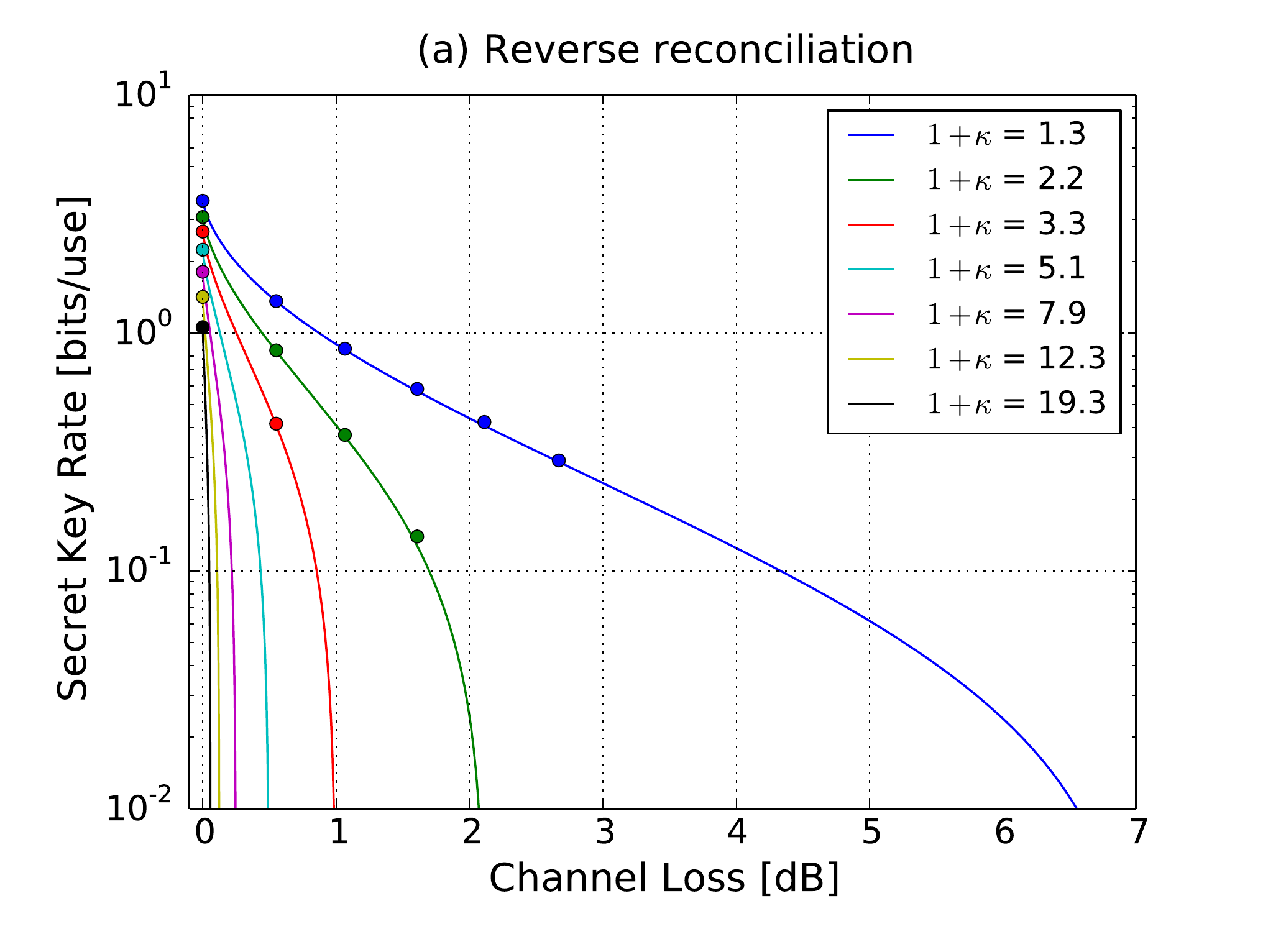}\label{fig:RRsecondRun}}
    \subfloat{\includegraphics[width=0.5\textwidth]{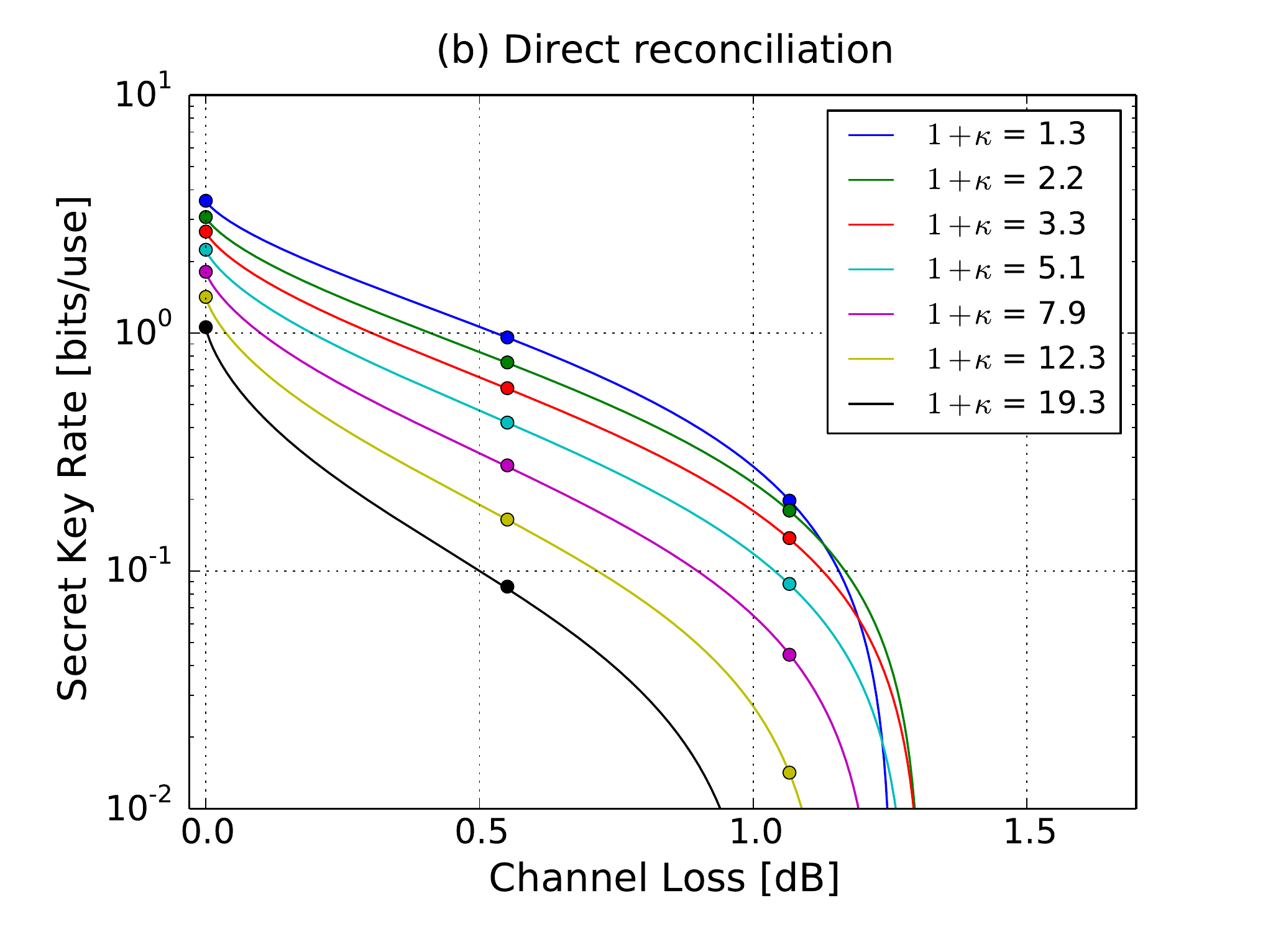}\label{fig:DRsecondRun}}
    \caption{Measured data and theory curves for different levels of preparation noise using ({\bf a}) reverse reconciliation and ({\bf b}) direct reconciliation in the post-processing. Error reconciliation efficiency $\beta = 95\,\%$. Due to our simulation of losses (see main text) the error bars on the channel loss are negligibly small and, thus, not shown in the figure.}
    \label{fig:Measurements}
\end{figure}

Figure~\ref{fig:Measurements} shows the experimentally obtained secret key rate for both reverse reconciliation and direct reconciliation for various levels of preparation noise $\kappa$. 
The modulation variance at Alice's station was set to about $32$ shot-noise units (SNU). 
The channel excess noise parameter $W$ was determined to $1.11$ SNUs.
To keep the shot-noise to electronic-noise clearance for all measurements constant and because the optical power in the two beams had to be the same to perform heterodyne detection, we simulated optical loss between Alice's and Bob's station.
The simulation was achieved by scaling the levels of the preparation noise and the signal modulations appropriately which is equivalent to transmission loss for coherent states.

From Figure~\ref{fig:RRsecondRun}, which shows the achieved secure key rate for reverse reconciliation, it is clearly visible that with increasing preparation noise not only does the secret key rate become smaller, so does the possible channel loss.
Especially for very high preparation noise the tolerable channel loss becomes very small. For the highest investigated preparation noise value of $1+\kappa \approx 19$ a channel loss of merely about $0.06$\,dB is possible.
This is in contrast to direct reconciliation, shown in Figure~\ref{fig:DRsecondRun}, where for this value of preparation noise about $0.85$\,dB channel loss can be reached.
While the strength of the direct reconciliation protocol lies in this region of high preparation noise, the theoretical behaviour seen in Figure~\ref{fig:Theory} of preparation noise enhanced tolerable channel loss is also resembled in the measurement.
For a preparation noise of $1+\kappa \approx 3$ the channel loss can be as large as $1.1$\,dB in comparison to about $1$\,dB without preparation noise.
This behaviour is, however, of purely theoretical interest since for these moderate values of preparation noise, reverse reconciliation still offers larger distances.


\section{Conclusions}

In conclusion, we have provided an experimental demonstrator for CV QKD with a noisy laser source. 
Using direct reconciliation we have shown that preparation noise of about $19$ SNUs can easily be tolerated, even though secret key rate and possible channel transmission become lower.
Theoretical calculations show that in fact the preparation noise can be arbitrarily high and still it should be possible to extract a secret key~\cite{Weedbrook2012}.

While in our experiment we simulated the effect of preparation noise in a controlled fashion, implementing the protocol with a noisy diode laser instead will demonstrate the protocol's ability to provide secret keys over short distances despite the noise of the source.


\acknowledgments{Acknowledgments}

This research was supported by the Danish Council for Independent Research, Technology and Production Sciences (Sapere Aude program). T.G.\ acknowledges support from the H.C.\,\O rsted postdoctoral programme.


\authorcontributions{Author Contributions}

All authors conceived the experiment. Christian S. Jacobsen implemented the setup under supervision of Tobias Gehring and Ulrik L. Andersen. Christian S. Jacobsen and Tobias Gehring performed the experiment, Christian S. Jacobsen analysed the experimental data. Christian S. Jacobsen and Tobias Gehring wrote the manuscript with contributions from Ulrik L. Andersen. All authors have read and approved the final manuscript.


\conflictofinterests{Conflicts of Interest}

The authors declare no conflict of interest.

\end{document}